\documentclass[%
 reprint,
unsortedaddress,
 amsmath,amssymb,
 aps,
 prl,
]{revtex4-1}

\usepackage{soul}
\usepackage{graphicx}% Include figure files
\usepackage{dcolumn}% Align table columns on decimal point
\usepackage{bm}% bold math
\usepackage[mathlines]{lineno}% Enable numbering of text and display math
\usepackage{xcolor}
\usepackage[normalem]{ulem}
\begin{document}

%\preprint{APS/123-QED}

\title{Observation of transient and asymptotic driven structural states of tungsten exposed to irradiation}% Force line breaks with \\

\author{Daniel R. Mason}%
\affiliation{UK Atomic Energy Authority, Culham Science Centre, Oxfordshire OX14 3DB, UK}%
\email{Daniel.Mason@ukaea.uk}

\author{Suchandrima Das}%
\address{Department of Engineering Science, University of Oxford, Parks Road, OX1 3PJ, UK}%
\email{Suchandrima.Das@eng.ox.ac.uk}%

\author{Peter M. Derlet}
\affiliation{Condensed Matter Theory Group, Paul Scherrer Institut, CH-5232 Villigen PSI, Switzerland}
\email{Peter.Derlet@psi.ch}

\author{Sergei L. Dudarev}
\affiliation{UK Atomic Energy Authority, Culham Science Centre, Oxfordshire OX14 3DB, UK}
\email{Sergei.Dudarev@ukaea.uk}
\author{Andrew London}
\affiliation{UK Atomic Energy Authority, Culham Science Centre, Oxfordshire OX14 3DB, UK}

\author{Hongbing Yu}
\affiliation{Department of Engineering Science, University of Oxford, Parks Road, OX1 3PJ, UK}

\author{Nicholas W. Phillips}
\affiliation{Department of Engineering Science, University of Oxford, Parks Road, OX1 3PJ, UK}

\author{David Yang}
\affiliation{Department of Engineering Science, University of Oxford, Parks Road, OX1 3PJ, UK}

\author{Kenichiro Mizohata}
\affiliation{University of Helsinki, P.O. Box 64, 00560 Helsinki, Finland}

\author{Ruqing Xu}
\affiliation{Advanced Photon Source, Argonne National Lab, 9700 South Cass Avenue, Argonne, IL 60439, USA}

\author{Felix Hofmann}
\affiliation{Department of Engineering Science, University of Oxford, Parks Road, OX1 3PJ, UK}
\email{felix.hofmann@eng.ox.ac.uk}

\begin{abstract}
Combining spatially resolved X-ray Laue diffraction with atomic-scale simulations, we observe how ion-irradiated tungsten undergoes a series of non-linear structural transformations with increasing irradiation exposure. Nanoscale defect-induced deformations accumulating above 0.02 displacements per atom (dpa) lead to highly fluctuating strains at $\sim$0.1 dpa, collapsing into a driven quasi-steady structural state above $\sim$1 dpa. The driven asymptotic state is characterized by finely dispersed vacancy defects coexisting with an extended dislocation network, and exhibits positive volumetric swelling due to the creation of new crystallographic planes through self-interstitial coalescence, but negative lattice strain.
\end{abstract}

\date{\today}                              
                              
\maketitle
Effects of irradiation on materials and their implications for structural integrity are major concerns for the design and operation of advanced nuclear power reactors \cite{Zinkle2013,Knaster2016}. Direct mechanistic models can correlate the evolution of irradiation-induced residual stresses and strains with components' lifetime \cite{marian2017recent,dudarev2018multi}, however the dynamics of the damage microstructure are complex and non-linear, span multiple length and time scales, and vary with exposure and environmental conditions \cite{Kiener2011,Was2017}. It remains challenging to account for contributing factors at relevant length- and time-scales with a minimum-parameter model.    

Quantitative experimental observations of irradiation effects require samples formed under controlled conditions of exposure, temperature, and applied stress. Ion-irradiation offers a cost- and time- effective alternative to neutron irradiation avoiding sample activation \cite{Was2012}, and real-space observations of microstructure produced by ion-irradiation have contributed extensively to the development of highly irradiation-resistant materials \cite{Silcox1959,SMALLMAN2014251,Zinkle2013}. Experimental techniques sensitive to the few-micron-thick ion damaged layer include transmission electron microscopy (TEM) \cite{Yi_Acta2015,Yi_Acta2016,Yi2015,ElAtwani_Acta2018, Ciupinski2013, Guo2020, Harrison2018, Ipatova2019}, X-ray diffraction \cite{Das2018,DeBroglie2015,Hofmann2015}, positron annihilation spectroscopy \cite{Debelle_JNM2008, Hu2016}, micro-mechanical tests \cite{Kiener2011,Armstrong2013a,Hwang2016, Hosemann2012} and laser-based techniques \cite{Short_JOM2015,Hofmann_SciRep2015, Dennett2019, Dennett2018, Reza2020}. 

Transferable interpretation of ion-irradiated materials data is an outstanding challenge. Quantitative models for irradiation effects are restricted to pure crystalline materials and very low exposure, $10^{-6}$ to $10^{-4}$ displacements per atom (dpa) \cite{Fu2005,Ortiz2007}. At high doses, consistent and unambiguous analysis proves difficult, and the  interpretation of experiments relies on temperature-dose rate scaling \cite{Was2012}, rate theory \cite{Jourdan2012} or cluster dynamics \cite{Marian_JNM2011,Dunn_JNM2013}. These models use kinetic equations involving potentially a multitude of parameters, and do not treat the microscopic fluctuating stresses and strains that drive defect interactions at the nano-scale \cite{Hirth_and_Lothe,Dudarev_PRB2010,Mason_JPCM2014}.

The spatial variation of strains and stresses observed in irradiated materials \cite{PHILLIPS2020219,yu2020nondestructive} can directly validate real-space simulations, since elasticity equations relate atomic-scale defects to macroscopic strains \cite{dudarev2018multi}. Here, we demonstrate this principle using an effectively parameter-free model to capture the physics of defect microstructure evolution without an over-reliance on thermal activation. The 3D depth-resolved lattice strain induced by the {\it entire} population of irradiation defects is probed with $\sim$ $10^{-4}$ strain sensitivity using synchrotron X-ray micro-beam Laue-diffraction, and interpreted quantitatively by direct atomic level simulations. The approach offers a unique advantage over TEM observations that only image defects larger than a critical size \cite{Zhou2006,Yi_Acta2015,Yi2016,Hwang2016}.

Tungsten, the front-runner candidate for armour components in ITER \cite{rieth2013recent,reiser2020}, serves as the prototype material for this study. In service, tungsten is anticipated to encounter significant radiation exposure \cite{Gilbert_2012}. The dose-dependent irradiation-induced defect microstructure in tungsten, under realistic operating conditions, is key to determining component lifetime and power plant availability. Currently, detailed qualitative information about microstructure is fragmented, particularly at ambient temperature for dense defect populations \cite{Kiener2011} where the mobility of defects is suppressed, resulting in exceedingly long relaxation times \cite{Ferroni_ActaMat2015,Papamihail2016}. Here, we show how the non-linear evolution of microstructure can be understood quantitatively by a systematic experimental and simulation study of ion-irradiated tungsten exposed to a wide range of doses at room temperature.

\section{Experimental Observations}

\begin{figure}[h!tb]
    \centering
    \includegraphics[width=0.9\linewidth]{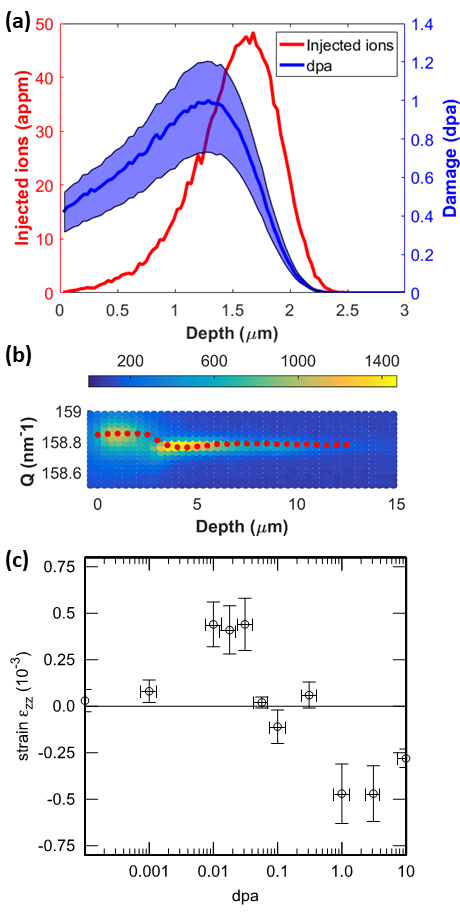} 
\caption{(a) Injected tungsten ion concentration and displacement damage, calculated using SRIM, for the 1 dpa sample. The blue solid line shows nominal dpa predicted using a threshold displacement energy of 68 eV. The shaded region shows upper and lower dpa bounds, corresponding to threshold displacement energies of 55 eV and 90 eV respectively. (b) Diffracted X-ray intensity, integrated in the tangential reciprocal space directions, for the (008) Bragg peak of the 1 dpa tungsten sample. Intensity is shown as a function of the scattering vector magnitude $|{\bf q}|$ and sample depth. The superimposed red dotted line shows the fitted peak centres $q_{\mathrm{fit}}(d)$. (c) Depth-averaged strain measured in ion implantation experiments. Horizontal error bars indicate the dpa uncertainty associated with the variation of assumed threshold displacement energies.}
    \label{fig:cra_analysis_expt}
\end{figure}

Tungsten samples were irradiated with self-ions to damage levels from 0.001 to 10 dpa. Details of sample preparation, ion-implantation method and fluences used are provided in the Appendix. Target displacements and ion ranges, estimated using the SRIM code \cite{ASTM,Ziegler2010}, show a $\sim$2.5 $\mu$m thick implanted layer (Fig. \ref{fig:cra_analysis_expt}(a)). 

Three $\langle$001$\rangle$ grains ($\sim$300 $\mu$m size) were identified in each implanted sample using electron back-scattering diffraction (EBSD). In each grain, the strain in the $\langle$001$\rangle$ direction was measured using depth-resolved Laue diffraction with $\sim$ $10^{-4}$ strain sensitivity \cite{Hofmann2015, Das2018Scripta, Das2018}. A polychromatic X-ray beam (7-30 keV) was focused to $\sim$300 nm FWHM using KB mirrors, and the sample placed at the beam focus in 45$^\circ$ reflection geometry. Diffraction patterns were recorded on an area detector $\sim$500 mm above the sample. A resolution of $\sim$500 nm along the incident beam direction was achieved using the differential aperture X-ray microscopy (DAXM) technique \cite{34ideHofmann2013, Liu2010, Larson2002, Das2018}.

A 3D reciprocal space map of each ($00n$) reflection was measured by monochromating the incident beam ($\Delta E/E\sim 10^{-4}$) and scanning the photon energy \cite{Chung1999, Das2018}. More information about the diffraction measurements is provided in the Appendix. Fig. \ref{fig:cra_analysis_expt}(b) shows the diffracted intensity, integrated over the tangential reciprocal space directions, plotted as a function of the scattering vector magnitude $|\bm{q}|$ and depth in the sample. The broad peak between 0 and $\sim$2.5 $\mu$m corresponds to the implanted layer, whereas the sharp peak at $\gg$2.5 $\mu$m corresponds to undamaged material. The measured implanted layer thickness is in good agreement with the SRIM prediction. 

Using the Laue data, we determine the lattice strain component normal to the sample surface. The peak centre $q_{\mathrm{fit}}(d)$ is found as a function of depth using the centre of mass method. In the small strain approximation, the lattice strain is then $\epsilon_{\mathrm{zz}}(d) = q_0/q_{\mathrm{fit}}(d) - 1$, where $q_0$ is the peak position for the reflection in an unstrained crystal, found here for each measurement using the average peak position in the last 1.5 $\mu$m depth (e.g. $d>11\mu$m in Fig. \ref{fig:cra_analysis_expt}(b)). 

To plot strain as a function of dose, we average the depth-dependent strain over the 2.5 $\mu$m implanted layer (Fig. \ref{fig:cra_analysis_expt}(c)). Strain in the 0.001 dpa sample is very small. At low fluence, between 0.01 and 0.032 dpa, lattice expansion is observed. A transition occurs between 0.056 and 0.32 dpa,  where the implantation-induced strains nearly vanish. At higher fluence ($>1$ dpa), we observe an apparent lattice contraction, manifested as negative lattice strain. This suggests a highly unusual dose-dependent change in the defect microstructure over the exposure interval spanned by the observations. We note that the dpa uncertainty associated with the choice of threshold displacement energy in SRIM calculations is small compared to the explored damage range (Fig. \ref{fig:cra_analysis_expt} (c)).

\section{Simulations and Interpretation}

To interpret experimental observations at the fundamental level of defect microstructure, we performed Frenkel Pair creation and relaxation simulations \cite{Debelle_JNM2008,Chartier_APL2016,Derlet_PRM2020} using the Creation Relaxation Algorithm (CRA) of Ref.~\cite{Derlet_PRM2020}. Each step of the algorithm randomly selects a number of atoms and randomly displaces them to new positions within the simulation cell. The structure is relaxed using LAMMPS \cite{LAMMPS} with an empirical potential for tungsten \cite{Mason_JPCM2017}, with zero stress condition in the $\hat{z}$-direction (oriented with [001]) and zero strain in the x-y-plane, reflecting the bulk constraint imposed by the substrate.

%\hl{The geometrical insertion Frenkel pairs is a drastic simplification of the experimentally occurring high-energy cascades (here due to 20 MeV tungsten ions} \cite{Sand_MRL2017,Sand_JNM2018} \hl{) but is generally found to predict qualitatively similar microstructures when compared to simulations of cascade evolution.} \cite{Derlet_PRM2020}.

\begin{figure}[h!tb]
    \centering
    \includegraphics[width=0.9\linewidth]{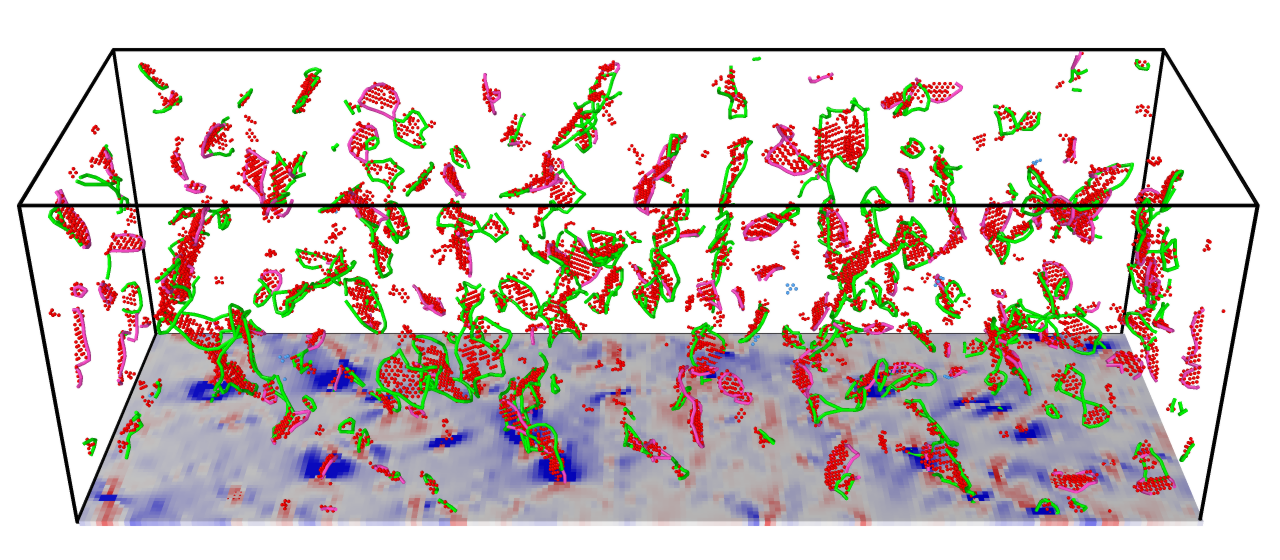}\\
    \includegraphics[width=0.9\linewidth]{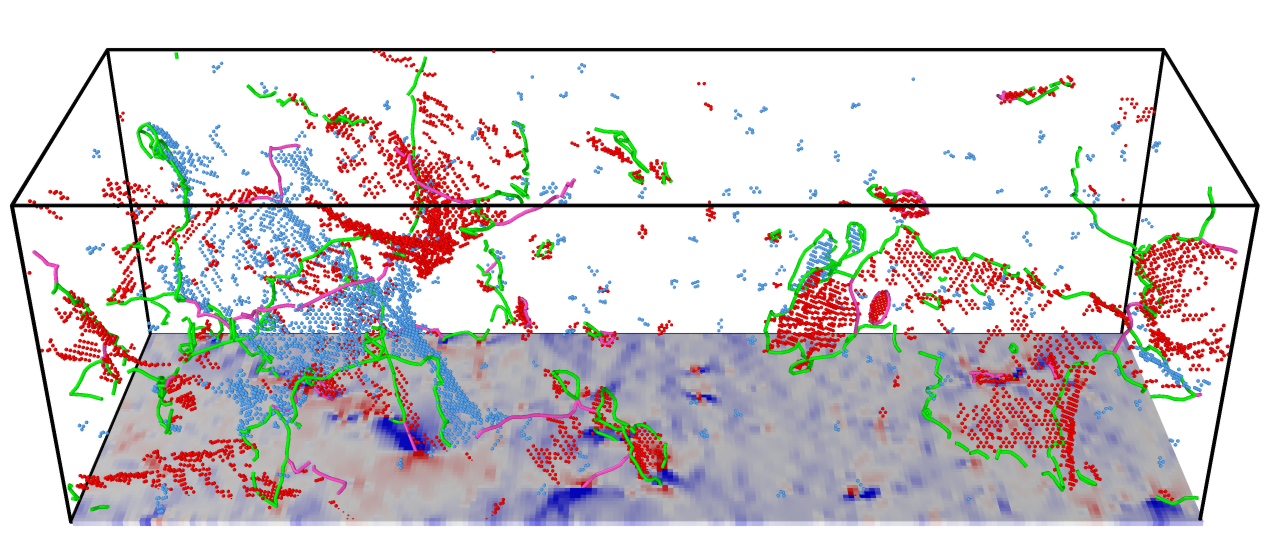}
    \caption{Representative Frenkel pair insertion simulations at 0.05 (top) and 0.3 (bottom) dpa using the CRA algorithm. The box size is $20.2\times20.2\times63.2$ nm$^3$, and the unconstrained cell dimension $\hat{z}$ is horizontal.   Vacancy (blue) and interstitial (red) clusters with $>3$ point defects are shown. Note the apparent formation of vacancy loops. A superimposed dislocation extraction algorithm (DXA) analysis \cite{Stukowski_MSMSE2012} shows both $1/2\langle 111\rangle$ (green) and $\langle 100\rangle$ (pink) dislocation lines. In the $y=0$ plane, the strain tensor component $\epsilon_{zz}$ is shown, with colour scale blue:white:red = -5\%:0:+5\% strain.}
    \label{fig:cra_simulations}
\end{figure}

This process is repeated many times and results in a microstructure that begins with isolated vacancy and intersitital defects and evolves, via interstitial dislocation loop nucleation and coalescence, to an extended dislocation network. The ratio of Frenkel pairs inserted to total atom content is the canonical dpa dose (cdpa) \cite{Derlet_PRM2020}. Representative results in Fig. \ref{fig:cra_simulations} show realizations of the microstructure at 0.05 cdpa and 0.3 cdpa. At 0.05 cdpa, the developing internal stress field has driven some of the interstitials to nucleate into dislocation loops, which by 0.3 cdpa have coalesced to extended dislocation structures, resulting in a microstructure that is insensitive to further Frenkel pair insertion~\cite{Derlet_PRM2020}. Additional information about the atomistic simulations can be found in the Appendix.  

Frenkel pair insertion is a drastic simplification of the 20 MeV self-ion cascades used in experiment\cite{Sand_MRL2017,Sand_JNM2018}, but predicts microstructures qualitatively similar to overlapping molecular dynamics cascade simulations
\cite{Derlet_PRM2020,Granberg_JNM2020}. It should be noted that there is no thermal activation in CRA simulations- all relaxation is stress driven- so CRA describes microstructures where long-range diffusion does not occur. For the present case of high purity, low temperature tungsten, vacancy migration is inactive \cite{NguyenManh_PRB2006}. The strong asymmetry in athermal mobility between vacancies and interstitials is therefore a justifiable physical limit and central to the observed simulated structural evolution. However, for materials that contain defect structures (impurities, sessile dislocation structures, etc.) that hinder interstitial mobility \cite{Bakaev_JAP2019,Mason_JPCM2014} and reduce this asymmetry, the situation is less clear but addressable using a combination of dedicated experiments and CRA simulations as done here.
 
\begin{figure}[h!tb]
    \centering
    \includegraphics[width=1.2\linewidth,angle=0]{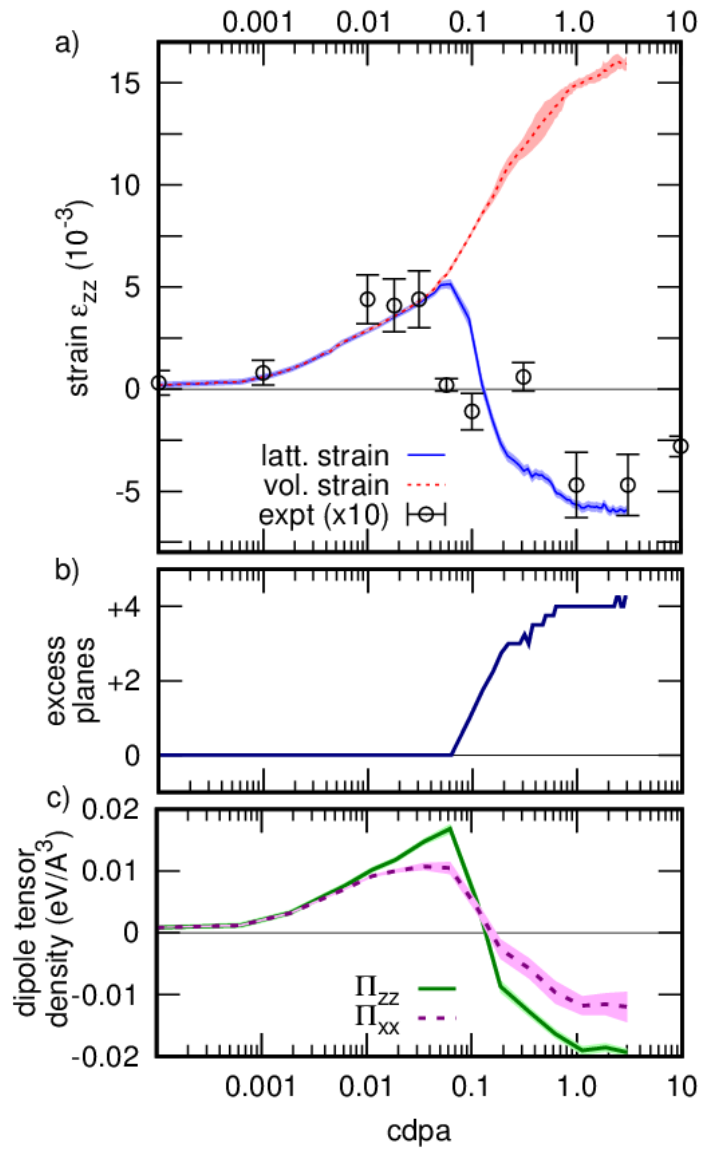} 
    \caption{(a) Lattice strain and volumetric strain (dashed) derived from simulations. Shaded region denotes one standard deviation. The experimental strain data is scaled by a factor of 10 to compare trends as well as absolute values. Volumetric strain due to the injected self-ions is small. (b) Number of excess planes recorded in the simulation. (c) Defect dipole tensor density (see text) computed from simulation cell stress. Note the horizontal scale is the same for all three plots.
      }
    \label{fig:cra_analysis_combi}
\end{figure}

 As in experiment, a measure of lattice strain can be obtained from a diffraction pattern, which for the case of simulation can be determined straightforwardly from the atomic positions of the microstructure produced by the Frenkel insertion method. Kinematic diffraction theory gives the diffraction spot intensity as being proportional to the square of the structure factor, $I(q)\propto \left| S(q) \right|^2$, where
    \begin{equation}
        S(q) = 1/\sqrt{N} \sum_j \exp\left[ i\, q z_j \right].
    \end{equation}
Here, both $q$ and $z_j$ are along the out-of-plane z-direction with the latter being the z-position of atom $j$. We use the simulated $[002]$ spot to find $q_{\mathrm{fit}}$ and hence the lattice strain as above. The resulting lattice strain is plotted in Fig. \ref{fig:cra_analysis_combi}a) as a function of cdpa and demonstrates similar behaviour to that seen in experiment, peaking at a cdpa of 0.05 after which it becomes negative at higher values of cdpa. Whilst there is remarkably good quantitative agreement as a function of dose, the scale of the simulated lattice strain is an order of magnitude larger than in experiments. This difference may be attributed to the absence of structural relaxation arising from thermal fluctuations \cite{marian2017recent}.

Fig. \ref{fig:cra_analysis_combi}a) also plots the volumetric strain associated with the change in volume of the simulation cell, defined as $\epsilon_{\mathrm{vol}} = L/L_0 - 1$.
Here $L$ is the evolving simulation cell periodic length along the z-direction. The volumetric strain initially follows the lattice strain, indicating that it arises directly from a homogeneous lattice expansion, which in this case is due to the low dose microstructural regime of lattice intersitial and vacancies. However at doses of approximately 0.05 cdpa, the volumetric strain decouples from the lattice strain and continues to increase with dose. 
%\hl{This corresponds to a microstructural regime in which interstitials cluster together to form dislocation loops that grow in size and eventually coalesce, resulting in the creation of new crystal planes. These new crystal planes are directly responsible for the observed increase in volume (along the $z$-direction), as can be seen in Fig. }\ref{fig:cra_analysis_combi}b) \hl{which displays the net number of new crystal planes forming as a function of dose. This observation agrees with reports in other materials of the lattice plane creation as a volumetric swelling mechanism} \cite{Brailsford1981}. 
In this regime, interstitials cluster to form dislocation loops that grow in size and eventually coalesce, resulting in the creation of new crystal planes along the $z$-direction seen in Fig. \ref{fig:cra_analysis_combi}b). This process preserves the increase in volume due to interstitial defects, while converting metastable microstructure into near-perfect crystal. This observation agrees with reports in other materials of lattice plane creation as a volumetric swelling mechanism \cite{Brailsford1981,Woo_PSS1990}. 
The good agreement between these simulations and experiment allows us to conclude that the change in the sign of lattice strain observed in experiment should {\it not} be interpreted as a transition from irradiation induced swelling to irradiation induced contraction.

%
%Such a decoupling between volumetric strain and lattice strain has been observed in past work focusing on bulk isotropic conditions~\cite{Derlet_PRM2020}, leading to a zero lattice strain at high values of cdpa. This earlier result can be rationalized by the strong internal stress fields driving the microstructure to a configuration that minimizes the overall energy of the system. The observation that the anisotropic conditions of fixed lateral strain and fixed out-of-plane zero-stress of the present work produce a microstructure that results in a negative out-of-plane strain at high doses needs further analysis.
Such a decoupling between volumetric and lattice strains has been used to infer vacancy concentrations in metals \cite{Simmons_PR1960}, and has also been observed in simulations under bulk isotropic conditions \cite{Derlet_PRM2020}. The latter lead to a zero lattice strain at high doses, whereas in the present case symmetry breaking leads to an asymptotic energy minimum with net negative out-of-plane strain. The general high dose strain condition as a function of sample boundary conditions, elastic constants, and defect densities needs further analysis.

 Using the elastic dipole tensor formalism to represent defects as sources of stress \cite{Leibfried}, and taking into account the zero $x,y$-strain condition imposed by the substrate and the traction-free condition at the surface, we find the non-vanishing components of lattice strain and stress in the irradiated layer
$\epsilon _{\mathrm{zz}}=(\Pi _{\mathrm{zz}}/2\mu)(1-2\nu)/ (1-\nu)$ and 
$\sigma _{\mathrm{xx}}=\sigma_{\mathrm{yy}}=\Pi_{\mathrm{xx}}-\nu \Pi _{\mathrm{zz}}/(1-\nu)$. Here $\Pi _{ij}$ is the volume density of dipole tensors of defects $\Pi _{ij}({\bf r})=\sum _{a}p^{(a)}_{ij}\delta ({\bf r}-{\bf R}_a)$, and $\mu$ and $\nu$ are the shear modulus and the Poisson ratio of tungsten. 

Computing $\epsilon_{\mathrm{zz}}$ and $\sigma_{\mathrm{xx}}$ from simulations, 
we find that the lattice strain sign change coincides with the observation, in the simulated diffraction pattern, of the start of formation of additional atomic planes parallel to the surface,  see Fig. \ref{fig:cra_analysis_combi}b. These planes, formed by the coalescence of interstitial dislocation loops, preserve the volumetric strain in the material, but by converting interstitial defect content into crystal planes reduce the lattice strain of the irradiated layer. This is confirmed by all the components of the dipole density tensor becoming negative in the high dose limit. 

%\hl{The removal of interstitials results in an excess of vacancies within the material, and it is this vacancy population that leads to the negative lattice strain observed at higher doses. From this perspective, the smaller magnitude of lattice strain observed in experiments compared to simulations is likely due to the thermally activated clustering of vacancies, an aspect that is not captured by the present zero temperature atomistic simulations}

The simulated microstructure beyond 1 dpa is dominated by network dislocations, a small number of dislocation loops of both interstitial and vacancy type, and a large number of excess vacancies. The vacancy population, totalling $2.5 \pm 0.1$\% lattice sites unoccupied, leads to the observed net negative lattice strain. The smaller magnitude lattice strain seen experimentally is likely due to thermally activated defect recombination, an aspect not captured by the present atomistic simulations.

\begin{figure}[h!tb]
    \centering
    \begin{minipage}{0.24\linewidth}
        \includegraphics[width=0.8\linewidth]{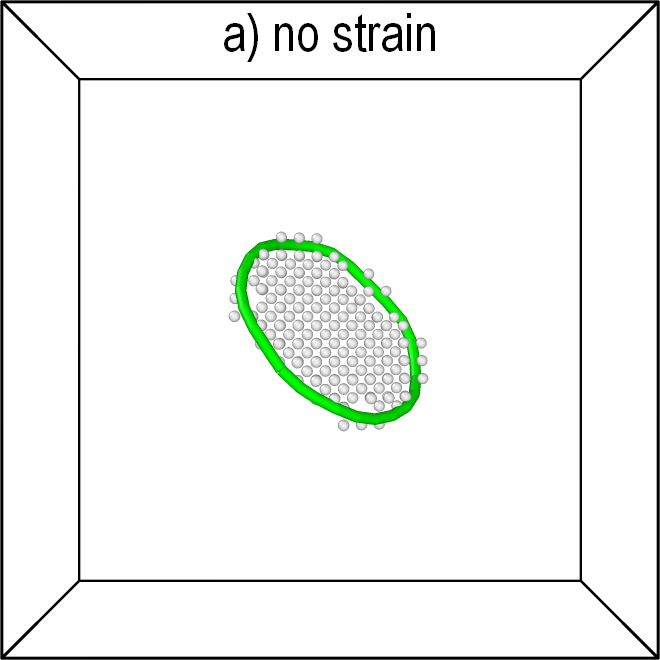}   \\
        \includegraphics[width=0.8\linewidth]{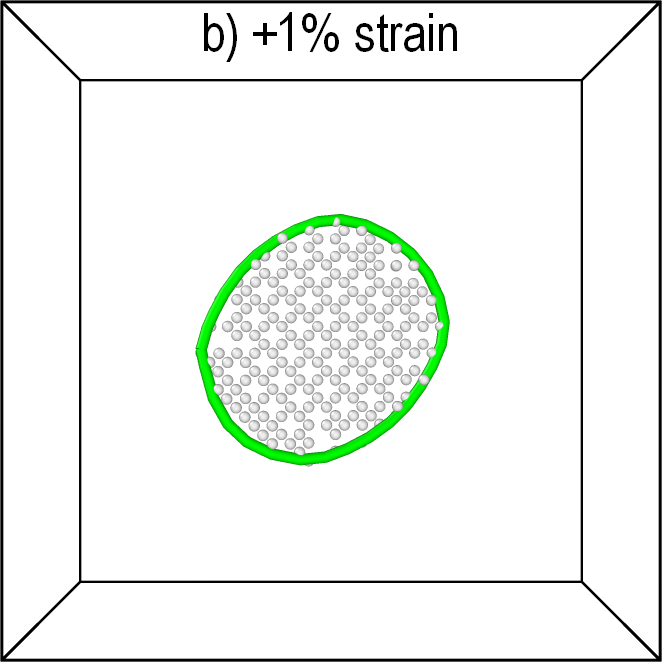}   \\
        \includegraphics[width=0.8\linewidth]{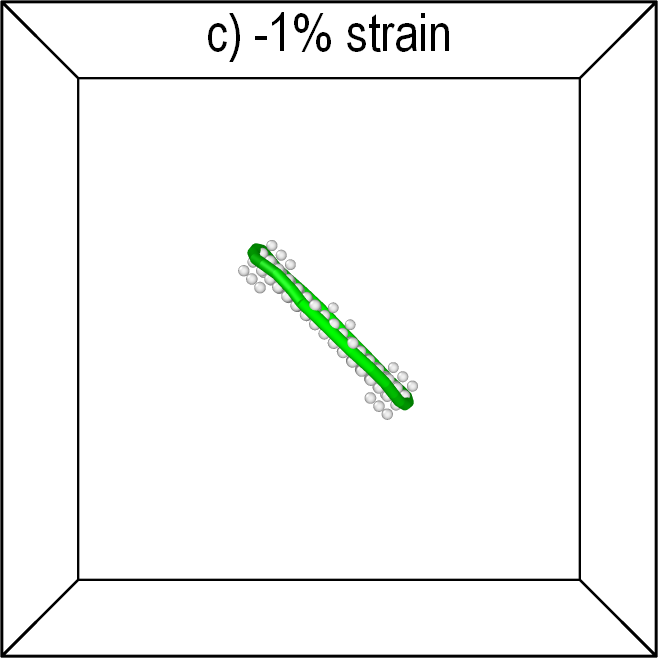}
    \end{minipage}
    \begin{minipage}{0.74\linewidth}
        \includegraphics[width=0.9\linewidth,angle=0]{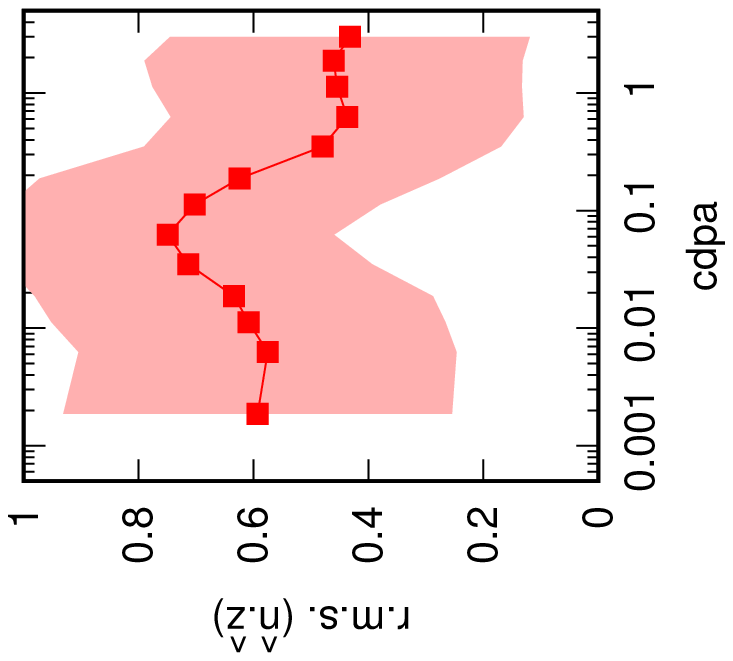} 
    \end{minipage}
    \caption{ Left: A 199-interstitial (4 nm diameter) glissile loop with $\mathbf{b}=1/2\langle 111\rangle$ relaxed under a small uniaxial strain (viewed along the strain axis) exhibits a spontaneous rotation of the habit plane with no change in $\mathbf{b}$. The effect is more pronounced for larger loops as the dipole tensor of a loop is proportional to the loop vector area \cite{Wolfer2004,Dudarev2018PRM}. In projection onto the Burgers vector direction, the change of orientation of the loop is undetectable. This stems from the conservation of the relaxation volume of the loop $({\bf b}\cdot {\bf A})$ \cite{Dudarev2018PRM},  where ${\bf A}$ is the vector area of the loop.   Right: root-mean-square orientation of habit plane normal as a function of dose. Shaded region is one standard deviation of the population. Standard error is order symbol size. }
    \label{fig:cluster_analysis_habitPlane_redo}
\end{figure}

The anisotropy of the dipole tensor density, emerging as a function of dose, is the result of self-action of the uniaxial stress field developing in the irradiated layer on the population of defects at a dose above $\sim 0.1$ dpa. The left panel of Fig. \ref{fig:cluster_analysis_habitPlane_redo}, a) b) and c) show how an isolated interstitial $\mathbf{b}=1/2\langle 111 \rangle$ dislocation loop changes its habit plane in response to an applied uniaxial strain. The response stems from the minimisation of energy of interaction of each individual defect with strain $E=-p_{\mathrm {zz}}\epsilon_{\mathrm {zz}}$, where $p_{\mathrm {zz}}$ is the $zz$ component of the dipole tensor of a defect, for example a dislocation loop \cite{Wolfer2004,Dudarev2018PRM}. The average orientation of the habit plane, $\mathbf {\hat{n}}$, of the interstitial loops and extended dislocation structures in our simulations is now measured and plotted via $\langle\mathbf{\hat{n}}\cdot\mathbf{\hat{z}}\rangle$ as a function of cdpa in the right panel of Fig. \ref{fig:cluster_analysis_habitPlane_redo}. This is done through numerically determining the optimal habit plane orientation of the dislocation structures identified by planes of interstitials. The figure reveals that at low dose this favours the orientation of habit planes of interstitial loops whose normals point in the out-of-plane direction, favouring the coalescence of loops into new atomic planes. On the other hand, in the high dose limit, where $\epsilon _{\mathrm {zz}} < 0$, the habit plane normal vectors of interstitial loops reorient tending now to point more towards the in-plane direction. As a result, no additional atomic crystal planes are formed beyond $\sim 0.6$ dpa. It is noted that such habit plane reorientation is a low barrier-energy process that occurs even under zero-loading conditions due to thermal fluctuations \cite{Osetsky2000,Derlet2011}. 

The negative lattice strain developing in the high dose limit is therefore a non-linear self-consistent phenomenon resulting from the interaction of radiation defects with the anisotropic uniaxial stress state developing in the irradiated layer. 

\section{Conclusions}

We find that upon ion irradiation of a tungsten surface layer, the measured out-of-plane lattice strain transitions from a positive to negative out-of-plane lattice strain. Through the use of the Creation Relaxation Algorithm atomistic simulation method, this behaviour is found to stem from the non-linear self-consistent interaction of radiation defect microstructure with its own stress field, due to constraint imposed by the un-implanted substrate material. The macroscopic volumetric strain, on the other hand, increases monotonically reflecting the well known phenomenon of irradiation induced swelling. The observed effect is likely to be a fundamental common feature of ion irradiation experiments, offering a simple and direct way of assessing the effect of stress and strain fields on defects produced in materials by irradiation. The present results also highlight that high dose irradiation can induce significant internal elastic loading, leading to dimensional changes and radiation-induced creep, all of which can adversely affect material components during operation of advanced fission and fusion reactors.

\section*{Acknowledgements} 
This work was funded by Leverhulme Trust Research Project Grant RPG-2016-190. X-ray diffraction experiments used resources of the Advanced Photon Source, a U.S. Department of Energy (DOE) Office of Science User Facility operated for the DOE Office of Science by Argonne National Laboratory under Contract No. DE-AC02-06CH11357. This work was carried out in part within the framework of the EUROfusion Consortium and has received funding from the Euratom research and training programme 2019-2020 under grant agreement No 633053 and from the RCUK Energy Programme [grant number EP/T012250/1]. We also acknowledge funding from the European Research Council (ERC) under the European Union’s Horizon 2020 research and innovation programme (grant agreement no. 714697). The views and opinions expressed herein do not necessarily reflect those of the European Commission. This work was also partly supported by the EPFL Swiss Plasma Center. Ion implantations were performed at the Helsinki Accelerator laboratory, Department of Physics, University of Helsinki. We would like to thank D. Perez, M. Boleininger, M.C. Marinica, M.J. Caturla and P.-W. Ma for stimulating discussions.

\section{appendix} 

\section{Details of sample preparation and ion-implantation}\label{appendix:ion-imp_table}

Eleven samples (10 $\times$ 10 $\times$ 1 mm$^3$) were cut from a polycrystalline tungsten sheet (procured from Plansee, nominal purity 99.99$\%$ by weight), fully recrystallised  at 1500 $^\circ$C for 24 hours in $~10^{-5}$ mbar vacuum. Samples were mechanically ground, polished with diamond paste and 0.1 $\mu$m colloidal silica, and electropolished in an electrolyte of 1$\%$ NaOH aqueous solution (8 V, 300 K) to obtain a mirror surface finish. 

Ten samples were implanted with 20 MeV tungsten ions at 300 K with a raster-scanned 5 mm diameter beam to obtain a spatially-uniform damage distribution. Irradiations used 20 MeV 184W (+5 charge state) ions with a 5 MV tandem accelerator \cite{TIKKANEN200435}. Raster scans were performed over a 15 $\times$ 15 mm$^2$ area using a sweeping frequency of 5-10Hz in both directions. Beam current and dose were monitored using a beam profilometer (BPM) before the target chamber. BPM current measurements were calibrated using a Faraday cup in the target chamber. A collimator (12.5mm diameter) was placed in front of the Faraday cup to define the area of the Faraday cup. The beam current was adjusted as a function of dose. Damage levels from 0.001 dpa to 1 dpa were exposed using beam current of 25 - 40 nA$/$cm$^2$, whilst the two highest doses were exposed using a beam current of about 90 nA$/$cm$^2$.

The ion doses required to reach a specific damage level were estimated using the SRIM code \cite{ASTM,Ziegler2010} (quick Kinchin-Pease model calculation). In the literature several different threshold displacement energies are recommended. Here, the nominal dpa dose corresponds to 68 eV threshold displacement energy (solid line in Fig. \ref{fig:cra_analysis_expt} (a)). An upper bound on dpa (55 eV threshold displacement energy \cite{Mason_JPCM2014}) and a lower bound (90 eV threshold displacement energy are also shown in Fig. \ref{fig:cra_analysis_expt} (a). The 90 eV threshold displacement energy is too high, however since it has been extensively used to calculate dpa in tungsten in previous publications, it is included for completeness. The ion fluence and corresponding damage rate used for each damage level are shown in Table \ref{fluence table}.

\begin {table}
\begin{center}
\begin{tabular}{ c c c }
  Nominal dose & Incident fluence & Damage rate\\ 
  (dpa)        &(ions/cm$^2$)     & (dpa/s) \\
  \hline
  0.001 & $2.42\times 10^{11}$ & 1.2 - 2.0 $\times 10^{-4}$   \\
  0.01  & $2.55\times 10^{12}$ & 1.2 - 2.0 $\times 10^{-4}$  \\  
  0.018 & $4.61\times 10^{12}$ & 1.2 - 2.0 $\times 10^{-4}$  \\
  0.032 & $8.2\times 10^{12}$  & 1.2 - 2.0 $\times 10^{-4}$  \\
  0.056 & $1.42\times 10^{13}$ & 1.2 - 2.0 $\times 10^{-4}$\\
  0.1   & $2.54\times 10^{13}$ & 1.2 - 2.0 $\times 10^{-4}$\\
  0.32  & $8.11\times 10^{13}$ & 1.2 - 2.0 $\times 10^{-4}$\\
  1.0   & $2.53\times 10^{14}$ & 1.2 - 2.0 $\times 10^{-4}$\\
  3.2   & $8.10\times 10^{14}$ & 4.4 $\times 10^{-4}$\\
  10.0  & $2.53\times 10^{15}$ & 4.4 $\times 10^{-4}$
 
\end{tabular}
\end{center}
\caption{Nominal damage level, corresponding 20 MeV tungsten ion fluence and damage rate for the considered tungsten samples.}
 \label{fluence table}
\end{table}

\section{Details of Laue diffraction}\label{appendix:Laue_details}
Laue measurements were performed at beamline 34-ID-E, Advanced Photon Source, Argonne National Laboratory, USA.
The order of the ($00n$) reflection, $n$, was chosen such that the diffraction peak centre was in the photon energy range of 17-22 keV. For each reflection, an energy interval of $\sim$80 eV was scanned with 2 eV steps. At each energy DAXM was also performed to resolve the depth dependence of the scattered intensity. Diffraction data was post-processed using the Laue-Go software package (J.Z.Tischler: tischler@anl.gov) and mapped into a 4D space volume defined by the reciprocal space coordinates ${q}_x$, ${q}_y$, ${q}_z$, and the distance along incident beam $d_{beam}$. 

\begin{figure}[!ht]
\centering
\includegraphics[width=7cm]{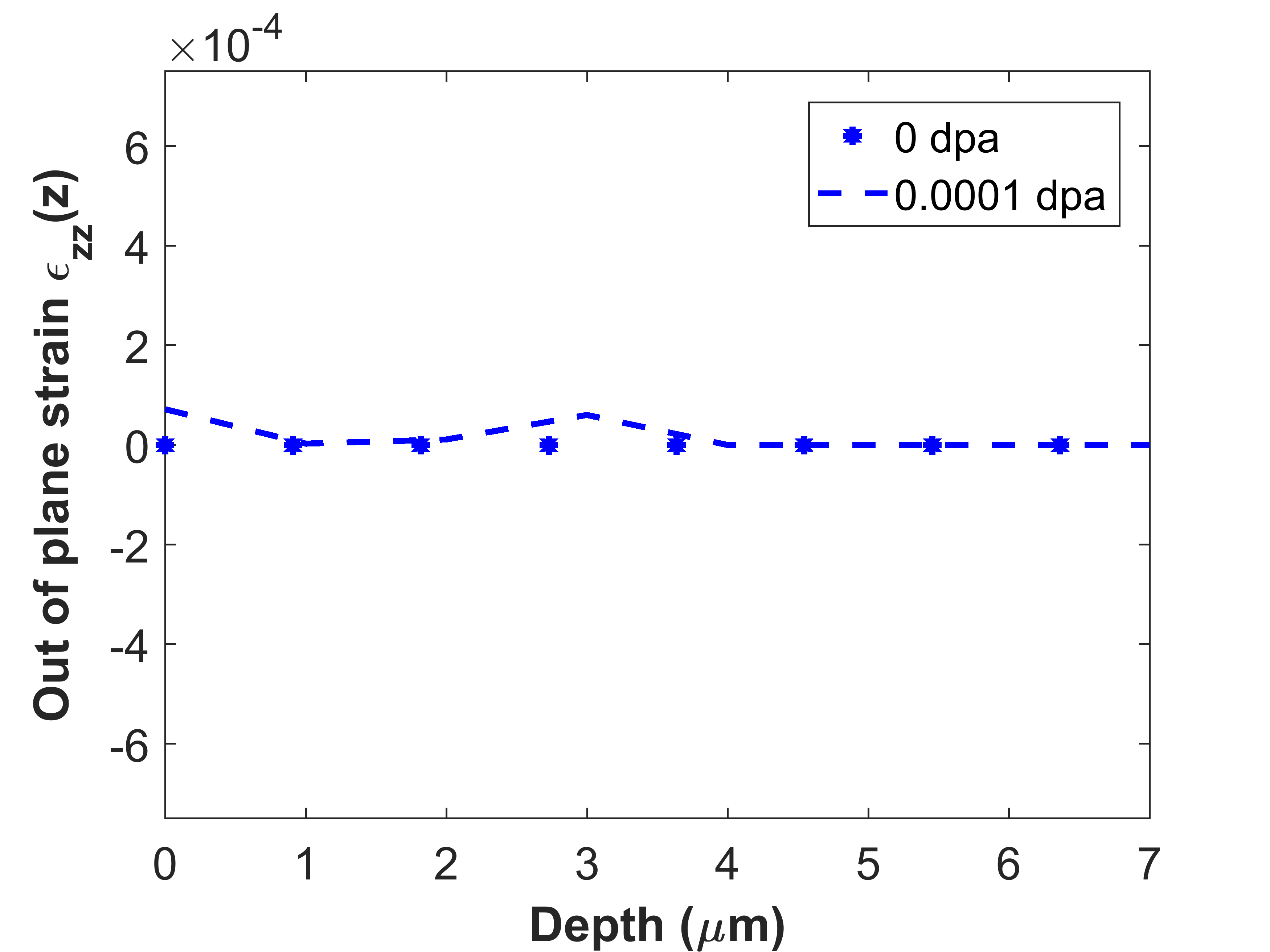}
\caption{Depth-resolved plot of $\epsilon_{zz}(z)$ for pure tungsten and 0.001 dpa self-ion implanted tungsten sample.}
\label{fig:unimp and 0_001 dpa strain curve}      
\end{figure}

Fig. \ref{fig:unimp and 0_001 dpa strain curve} shows the depth resolved plot of $\epsilon_{zz}(z)$ for pure tungsten and the 0.001 dpa self-ion implanted tungsten sample. It is seen that at 0.001 dpa, the implantation-induced strain is negligible. 

\begin{figure}[!ht]
\centering
\includegraphics[width=7 cm]{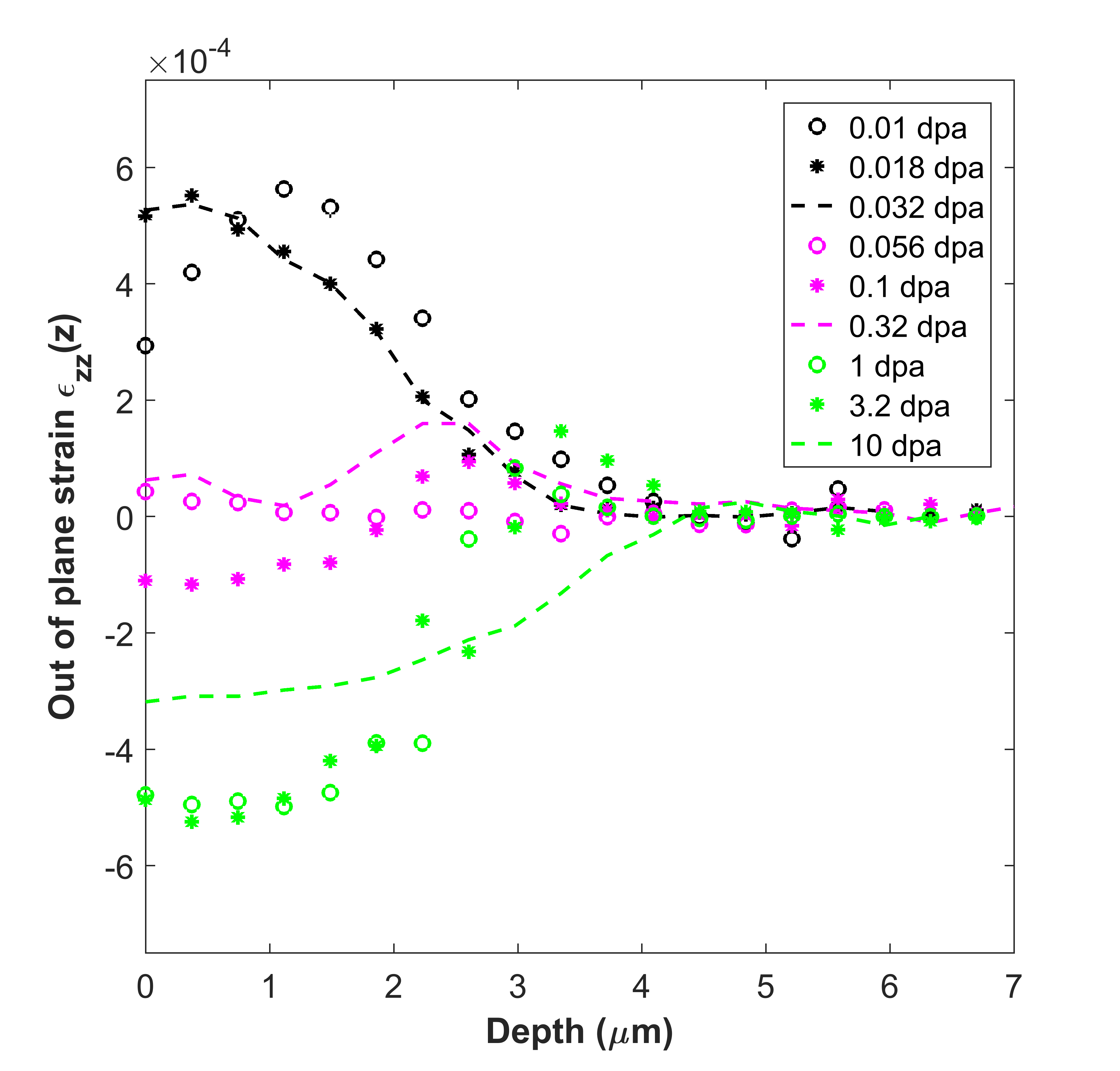}
\caption{Depth-resolved plot of $\epsilon_{zz}(z)$ for self-ion implanted tungsten samples of different damage levels.} 
\label{fig:self-ion_strain_avg_strain}      
\end{figure}

Fig. \ref{fig:self-ion_strain_avg_strain} shows $\epsilon_{zz}$ plotted as a function of depth in the sample for the self-ion implanted tungsten samples exposed to nine different damage levels $\geq 0.01$dpa. The curves in Fig. \ref{fig:self-ion_strain_avg_strain} are the average of three measurements for each sample. We note that although $\epsilon_{zz}(z)$ at the surface should vanish in agreement with the traction free boundary condition, this is not captured in Fig. \ref{fig:self-ion_strain_avg_strain} as our experiments integrate over a volume $\sim$ 500 nm cubed. Defects within this volume induce the strains still seen at depth marked 0 in Fig. \ref{fig:self-ion_strain_avg_strain}. 

\begin{figure}[!ht]
\centering
\includegraphics[width=7.5cm]{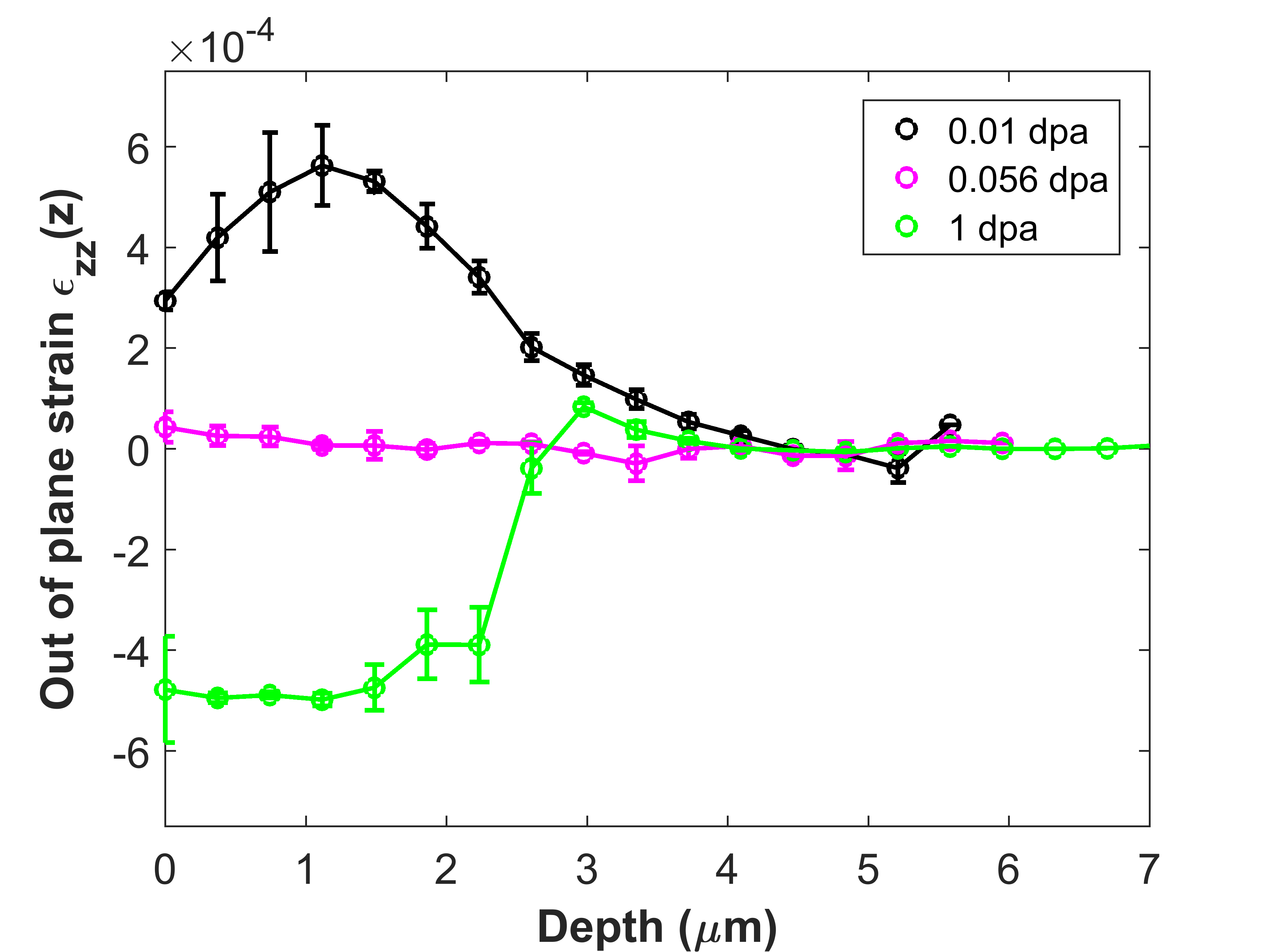}
\caption{Errorbars show $\pm$1 standard deviation of $\epsilon_{zz}(z)$ measurements at each depth for self-ion implanted tungsten samples exposed to 0.01, 0.056 and 1 dpa.} 
\label{self-ion strain 3 doses with errorbars set 1}
\end{figure}

\begin{figure}[!ht]
\centering
\includegraphics[width=7.5cm]{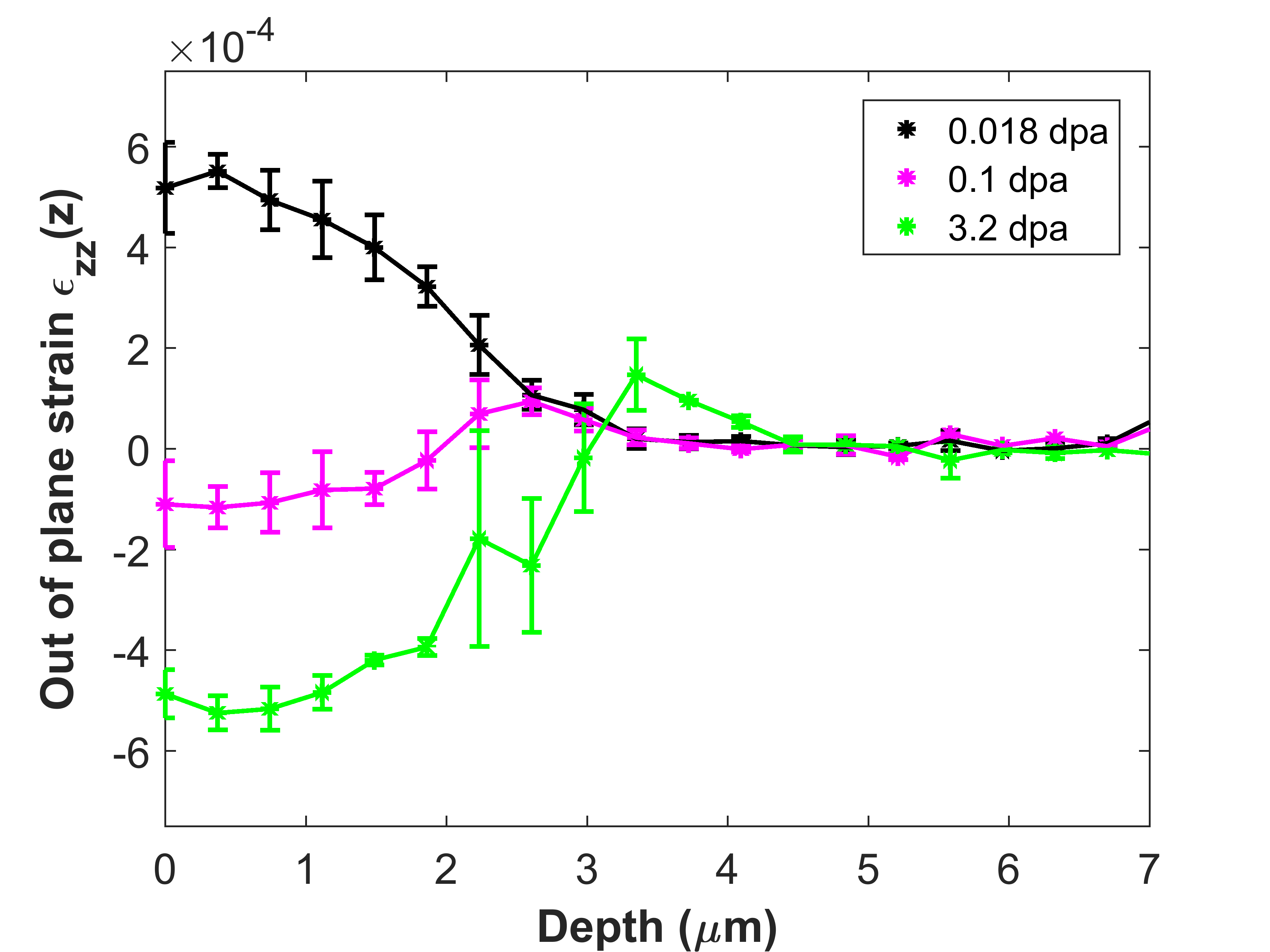}
\caption{Errorbars show $\pm$1 standard deviation of $\epsilon_{zz}(z)$ measurements at each depth for self-ion implanted tungsten samples exposed to 0.018, 0.1 and 3.2 dpa.} 
\label{self-ion strain 3 doses with errorbars set 2}
\end{figure}

\begin{figure}[!ht]
\centering
\includegraphics[width=7.5cm]{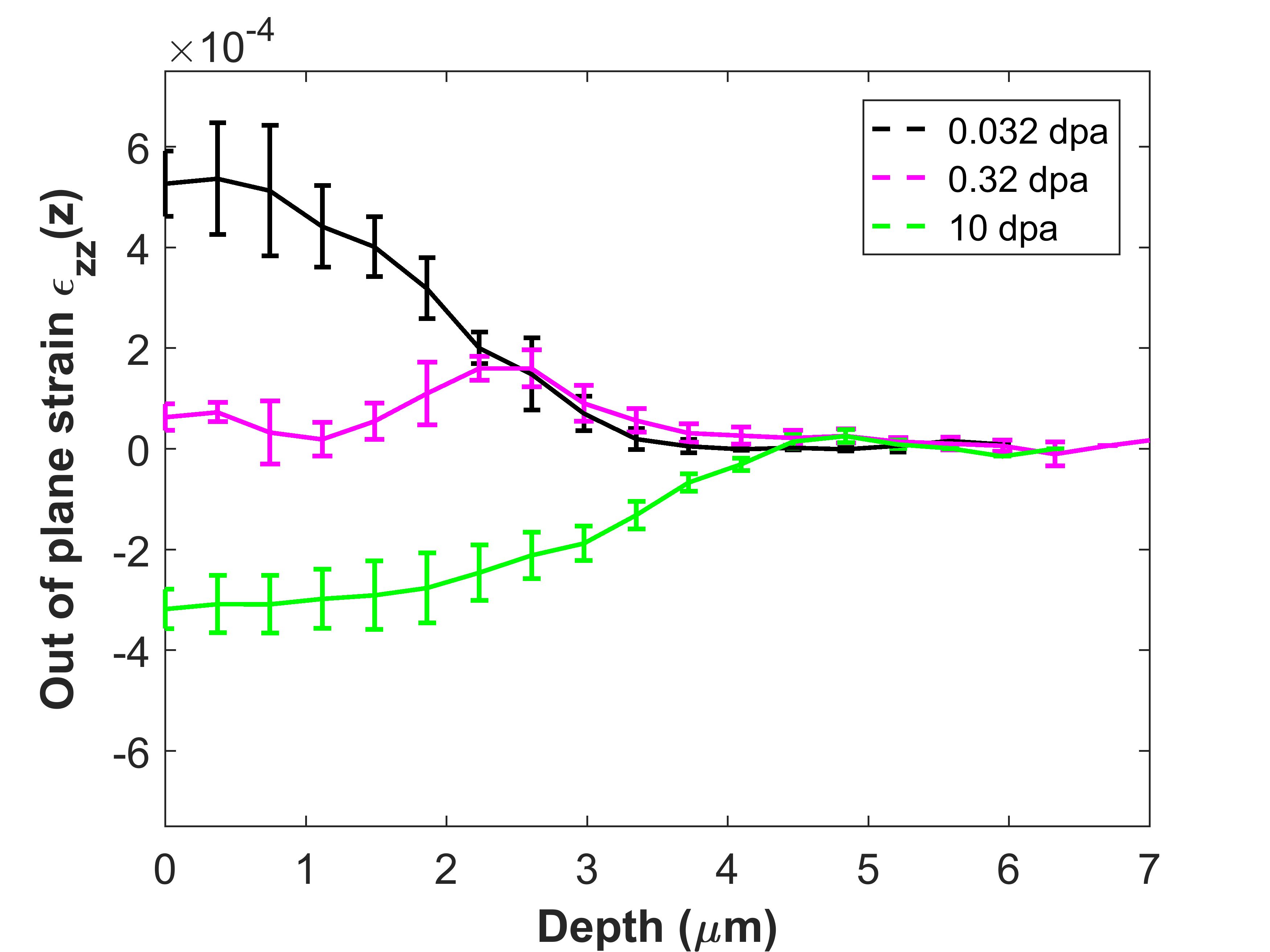}
\caption{Errorbars show $\pm$1 standard deviation of $\epsilon_{zz}(z)$ measurements at each depth for self-ion implanted tungsten samples exposed to 0.032, 0.32 dpa.} 
\label{self-ion strain 3 doses with errorbars set 3}
\end{figure}

For ease of visualisation, the errorbars showing $\pm$1 standard deviation across the multiple measurements for each sample are shown in three different plots in Fig. \ref{self-ion strain 3 doses with errorbars set 1}, Fig. \ref{self-ion strain 3 doses with errorbars set 2} and Fig. \ref{self-ion strain 3 doses with errorbars set 3}. 

\section{Details of atomistic simulation}
\label{appendix:atomic_simulation}

The used Frenkel insertion method, begins with a perfect BCC crystal. In the present work an insertion iteration involves randomly selecting $N$ atoms and randomly displacing them to positions elsewhere within the simulation cell. The atomic configuration is then relaxed to a new local potential energy minimum via the conjugate gradient method. In the work of Ref.~\cite{Derlet_PRM2020} $N=1$, whereas in the present work $N=1000$ (corresponding to 0.000625 cdpa per relaxation step). Using this larger value is computationally more efficient and results in microstructures whose characteristics are insensitive to the choice of $N<1000$. For the present work, a simulation cell of $64\times64\times200$ unit cells (1.6M atoms) was needed for convergence with respect to simulation cell size. Here the $x-y$ plane is the in-plane of the thin-film geometry and the $z$ plane is the out-of-plane direction. The conjugate gradient relaxation was performed under fixed in-plane zero strain and fixed out-of-plane zero stress conditions to correctly represent the thin-film boundary conditions. Introducing an explicit surface into simulations did not affect the main results of the work, indicating the observed strain phenomenon is due to a bulk anisotropy in the boundary conditions and not due to loop loss at a free surface. The presented results are obtained from four independent simulations.

The tungsten embedded atom method potential used (Ref. \cite{Mason_JPCM2017}) was chosen because it is known to produce good relaxation volumes for irradiation defects \cite{Mason_JAP2019} and therefore a correspondingly accurate far-field strain signature.

Since the number of lattice planes in a simulation can vary as a function of dose if interstitial loops present as new atomic planes, a robust method is needed to determine their number. The present work uses the best fit number of lattice planes, $n_z$, which is the value giving the maximum intensity in the simulated diffraction pattern given the known periodic cell length $L$, i.e.
	\begin{equation}
	    \label{eqn:commensurate_strain}
	    n_z = \mathrm{arg max} \left[ I(q = 4 \pi n_z / L) \right].
	\end{equation}

\end{document}